\documentclass[aps,showpacs]{revtex4}

\usepackage{graphicx}

\begin{document}

\title{Vacuum solutions  of five dimensional Einstein equations generated by inverse scattering method }

\author{Shinya Tomizawa, Yoshiyuki Morisawa and Yukinori Yasui}

\affiliation{ Department of Mathematics and Physics, Graduate School of Science, Osaka City University, 3-3-138 Sugimoto, Sumiyoshi,
Osaka 152-8551, Japan}

\date{\today}

\begin{abstract}
We study stationary and axially symmetric two solitonic solutions of five dimensional vacuum Einstein equations by using the inverse scattering method developed by Belinski and Zakharov. In this generation of the solutions, we use five dimensional Minkowski spacetime as a seed. It is shown that if we restrict ourselves to the case of one angular momentum component, the generated solution coincides with a black ring solution with a rotating two sphere which was found by Mishima and Iguchi recently. 
\end{abstract}


\maketitle

\section{Introduction}

In four dimensional Einstein-Maxwell systems, stationary black holes are considerably interesting objects since we would expect that any black holes formed by gravitational collapse settle down to a stationary state. Remarkably, stationary black holes are characterized only by total mass, angular momentum and electric (magnetic) charge. This fact is well known as {\it uniqueness theorems} of black holes~\cite{uniqueness}. Therefore, a lot of authors have investigated global and local properties of stationary black holes as the final state of black hole spacetimes.

Recently, higher dimensional black holes have much attention since the 
possibility of higher dimensional black hole production in a linear 
collider is predicted in TeV gravity~\cite{BHinCollider}. However, it is not clear whether black holes will be formed through the collisions of
protons in the linear collider. 
Though at first glance it seems that we cannot expect to obtain a complete description of real black holes due to the complexity of Einstein equations, essentially their nonlinearity, the uniqueness theorem of higher dimensional black holes would let us know the answer, at least, some information on the final states.  However, it is unlikely that in a higher dimensional stationary black hole spacetime there exists the uniqueness theorem in the sense of one in a four dimensional black hole spacetime, which means that the set of the parameters of mass, angular momentum and charge fails to determine a higher dimensional black hole uniquely. In fact, after the discovery of Myers-Perry solution
whose horizon is topologically $S^3$~\cite{Myers:1986un},  Emparan and Reall found a five dimensional rotating black ring solution with the
horizon homeomorphic to $S^1\times S^2$~\cite{Emparan:2001wn}. It has been found that there is a range within which these two solutions have the same mass and angular momentum. This implies the absence of the original uniqueness theorem like that of four dimensional black holes.  If one restricts the topology of a horizon to $S^3$, it has been shown that the only asymptotically flat black hole solution is Myers-Perry solution~\cite{Morisawa:2004tc}. We also comment that in Einstein-Maxwell-Chern-Simon theory, five dimensional black holes with spherical topology $S^3$ cannot be characterized by only these three physical parameters \cite{Kunz}, neither can black ring solutions coupled with form fields \cite{Emparan}.
In this view, what kind of black hole solutions there can exist in five 
or higher dimensional spacetimes is an interesting problem, and in order to 
find such black hole solutions admitted as possible, we have to develop 
generating-techniques of solutions of  higher dimensional Einstein equations.

In four dimensional Einstein gravity, systematic 
generation-techniques of stationary and axisymmetric solutions have been 
developed by a lot of authors~\cite{exact}. Recently, Mishima and Iguchi~\cite{Mishima:2005id} applied one of these techniques, Castejon-Amenedo-Manko's 
method~\cite{Castejon-Amenedo:1990b}, to five dimensional vacuum spacetimes and derived the 
asymptotically flat black ring solution which rotates in the azimuthal 
direction of two sphere. This solution is generated from five dimensional 
Minkowski spacetime as a seed. The expression of this solution  \cite{Private} as a C-metric coordinate was studied by Figueras~\cite{Figueras}. As a solitonic generation-technique in 
four dimensional Einstein gravity,
Belinski and Zakharov developed the inverse scattering method~\cite{Belinskii}. 
They generated Kerr-NUT solutions as simple two solitonic solutions on the 
Minkowski background. This technique is also used for the 
generations of multi-black hole solutions~\cite{exact}. We would expect 
that the application of this method to more than four dimensions might 
lead to the generation of a lot of physical solutions.
In fact, the static black ring solution \cite{Koikawa}, the five dimensional Schwartzschild solution \cite{Koikawa} and the Myers-Perry solution with two angular momentum components were reproduced 
by using this technique~\cite{Pomeransky:2005sj}.

In this article, we will apply the inverse scattering method established 
by Belinski et al. to five dimensional stationary and axisymmetric 
spacetimes, which means that there are a timelike Killing vector field and 
two axial Killing vector fields. We choose five dimensional Minkowski 
spacetime as a seed. Mishima-Iguchi's method gives solutions with only 
one angular momentum.
On the other hand, in general, this inverse scattering 
method generates solutions with two angular momentum components.
However, such a solution generated from Minkowski seed is not regular.
Therefore, in this 
article, we focus on the case with a single angular momentum and show that 
it coincides with a black ring solution with a rotating two sphere~\cite{Mishima:2005id,Pomeransky:2005sj}.

This article is organized as follows. In Sec.~\ref{sec:general}, we will  
review the result of Belinski's studies, and mention the 
normalization when we construct a physical metric, 
which satisfies the supplementary condition ${\rm det} g=-\rho^2$, from 
an unphysical metric. In Sec.~\ref{sec:Minkowski_seed}, we  will 
study  two solitonic solutions from five dimensional Minkowski spacetime as a 
seed. The solution with one angular momentum coincides with a black ring solution with a rotating two sphere~\cite{Mishima:2005id,Pomeransky:2005sj}. We summarize this article and give a discussion of related matters in Sec.~\ref{sec:summary}.

\section{$D$ dimensional $n$ soliton solutions}
\label{sec:general}
In this section, we review the results of Belinski's method~\cite{Belinskii} applied to  gravitational fields in general dimensions.
We begin with  $D$ dimensional stationary vacuum spacetimes with $D-2$ commuting Killing vector fields $(\partial/\partial t),(\partial/\partial x^{2}),\cdots, (\partial/\partial x^{D-2})$, where $(\partial/\partial t)$ is the Killing vector field associated with time translation and $(\partial/\partial x^2),\cdots,(\partial/\partial x^{D-2})$ denote the spatial Killing vectors. From the discussion in Ref.~\cite{Harmark:2004rm}, in vacuum spacetimes, the two-plane orthogonal to these Killing vector fields is integrable. 
In such spacetimes, the metric can be written in the canonical form~\cite{Harmark:2004rm} :

\begin{eqnarray}
ds^2=f(d\rho^2+dz^2)+g_{ab}dx^adx^b,
\end{eqnarray}
where $f=f(\rho,z)$ and $g_{ab}=g_{ab}(\rho,z) (a,b=1,\cdots,D-2)$ are 
a function and an induced metric on the $(D-2)$ dimensional plane, respectively.
Both of them depend only on $\rho$ and $z$.
The $(D-2)\times(D-2)$ matrix $g=(g_{ab})$ satisfies the constraint,  ${\rm det}\ g=-\rho^2$. From the 
vacuum Einstein equation $R_{ab}=0$, the matrix 
$g$ also satisfies the solitonic equation,

\begin{eqnarray}
(\rho g_{,\rho} g^{-1})_{,\rho}+(\rho g_{,z}g^{-1})_{,z}=0,\label{eq:soliton}
\end{eqnarray}
and from the other equations $R_{\rho\rho}+R_{zz}=0$ and $R_{\rho z}=0$, we obtain the equations which determine the function $f(\rho,z)$ for a given solution of the solitonic equation (\ref{eq:soliton}),
\begin{eqnarray}
(\ln f)_{,\rho}=-\frac{1}{\rho}+\frac{1}{4\rho}{\rm Tr}(U^2-V^2),\label{eq:f1}
\end{eqnarray}

\begin{eqnarray}
(\ln f)_{,z}=\frac{1}{2\rho}{\rm Tr}(UV),\label{eq:f2}
\end{eqnarray}
where the $(D-2)\times(D-2)$ matrices $U(\rho,z)$ and $V(\rho,z)$ are defined as\begin{eqnarray}
U=\rho g_{,\rho}g^{-1},\quad V=\rho g_{,z}g^{-1}.
\end{eqnarray}
The integrability condition with respect to $f$ is automatically 
satisfied for the solution $g$ of the equation~(\ref{eq:soliton}).
Then, the $n$ solitonic solution, which satisfies the constraint ${\rm det}\ g_{ab}=-\rho^2$, can be written as follows,

\begin{eqnarray}
g^{{\rm (phys)}}&=&(-1)^{\frac{n}{D-2}}\rho^{-\frac{2n}{D-2}}(\prod_{k=1}^n\mu_k^{\frac{2}{D-2}})g^{{\rm (unphys)}},\label{eq:nor}
\end{eqnarray}
where the unphysical metric $g^{{\rm (unphys)}}$, which is a solution of the equation (\ref{eq:soliton}) but does not meet the supplementary condition ${\rm det}\ g_{ab}=-\rho^2$, is given by

\begin{eqnarray}
g^{{\rm (unphys)}}=g_0-\sum_{k,l=1}^n\mu_k^{-1}\mu_l^{-1}\Pi^{kl}m^{(l)}_{0e}m^{(k)}_{0f}(g_0)_{ca}(g_0)_{db}[\psi_0^{-1}(\mu_l,\rho,z)]^{ec}[\psi_0^{-1}(\mu_k,\rho,z)]^{fd}\label{eq:unphys1}.
\end{eqnarray}
Here $g_0$ is an arbitrary seed solution 
and the poles $\mu_k$ are given by 
$\mu_k=w_k-z\pm\sqrt{(w_k-z)^2+\rho^2}$ 
together with arbitrary constants $w_k (k=1,\cdots,n)$ 
and $m_{0a}^{(k)}\ (k=1,\cdots,n; a=1,\cdots,D-2)$.
The matrix $\Pi^{kl}$ 
is the inverse of $\Gamma_{kl}$, which is given by 
$\Gamma_{kl}:=(\rho^2+\mu_k\mu_l)^{-1}m^{(k)}_{0c}[\psi_0^{-1}(\mu_k,\rho,z)]^{ca}m^{(l)}_{0b}[\psi_0^{-1}(\mu_l,\rho,z)]^{bd}(g_{0})_{ad}$. 
The generating matrix $\psi_0(\lambda,\rho,z)$ is a solution of the 
following equations

\begin{eqnarray}
D_1\psi_0=\frac{\rho V_0-\lambda U_0}{\lambda^2+\rho^2}\psi_0,\label{eq:psi1}
\end{eqnarray}
and
\begin{eqnarray}
D_2\psi_0=\frac{\rho U_0+\lambda V_0}{\lambda^2+\rho^2}\psi_0,\label{eq:psi2}
\end{eqnarray}
where  $\lambda$ is a complex parameter independent of the coordinates 
$\rho$ and $z$, and $U_0=\rho g_{0,\rho}g_0^{-1}$ and $V_0=\rho g_{0,z}g^{-1}_0$. 
The derivative operators $D_1$ and $D_2$ are defined by

\begin{eqnarray}
D_1=\partial_z-\frac{2\lambda^2}{\lambda^2+\rho^2}\partial_\lambda,\quad D_2=\partial_\rho+\frac{2\lambda\rho}{\lambda^2+\rho^2}\partial_\lambda.
\end{eqnarray}
Substituting the physical metric solution 
$g^{{\rm (phys)}}$ given by the equation (\ref{eq:nor}) into the equations 
(\ref{eq:f1}) and  (\ref{eq:f2}), we obtain a physical value of $f$ :

\begin{eqnarray}
f=Cf_0\rho^{-\frac{n(n+4-D)}{D-2}}{\rm det} (\Gamma_{kl})\prod_{k=1}^n[\mu_k^{\frac{2(n-3+D)}{D-2}}(\mu_k^2+\rho^2)^{\frac{4-D}{D-2}}]\cdot\prod_{k>l}^n(\mu_k-\mu_l)^{\frac{4}{2-D}},
\end{eqnarray}
where $C$ is an arbitrary constant, and $f_0$ is a value of $f$ 
corresponding to the seed $g_0$.

\subsection{five dimensional two solitonic solution}

As mentioned in the introduction,
we are interested in two solitonic solutions in five dimensional vacuum spacetimes as the simplest case. Therefore, let us put $D=5,\ n=2$ in the equations (\ref{eq:nor}) and (\ref{eq:unphys1}).  
Two soliton solutions are expressed in the form
\begin{eqnarray}
g_{ab}^{{\rm (phys)}}=\rho^{-\frac{4}{3}}(\mu_1\mu_2)^{\frac{2}{3}}g_{ab}^{{\rm (unphys)}}\label{eq:5sol},
\end{eqnarray}
where $g_{ab}^{{\rm (unphys)}}$ is a three dimensional unphysical metric which takes the form of

\begin{eqnarray}
g^{{\rm (unphys)}}=g_0-\sum_{k,l=1}^{2}\mu_k^{-1}\mu_l^{-1}\Pi^{kl}m^{(l)}_{0e}m^{(k)}_{0f}(g_0)_{ca}(g_0)_{db}[\psi_0^{-1}(\mu_l,\rho,z)]^{ec}[\psi_0^{-1}(\mu_k,\rho,z)]^{fd}\label{eq:unphys}.
\end{eqnarray}

For simplicity of notation, we hereafter put $m_{01}^{(1)}=a,\ m_{01}^{(2)}=b,\ m_{02}^{(1)}=c,\ m_{02}^{(2)}=d, \ m_{03}^{(1)}=e,\ m_{03}^{(2)}=f$.
The two poles are given by 

\begin{eqnarray}
\mu_1&=&w_1-z\pm\sqrt{(w_1-z)^2+\rho^2},\nonumber\\
\mu_2&=&w_2-z\pm\sqrt{(w_2-z)^2+\rho^2}.
\end{eqnarray}
Through the below, we put $w_1=-w_2=-\sigma$ and choose both of signs as $+$:
\begin{eqnarray}
\mu_1&=&-\sigma-z+\sqrt{(\sigma+z)^2+\rho^2}\label{eq:mu1},\\
\mu_2&=&+\sigma-z+\sqrt{(\sigma-z)^2+\rho^2}\label{eq:mu2}.
\end{eqnarray}

Here we comment that in general cases with two angular momentum components, the  solution in general is not regular on a part of an axis if the seed solution is regular there.  In fact, in a static limit (as such an example we can choose $a=b=c=d=0, e\not=0,f\not=0$), the solution becomes 

\begin{eqnarray}
g^{{\rm (phys)}}={\rm diag} \Biggl(\biggl(\frac{\mu_1\mu_2}{\rho^2}\biggr)^{\frac{2}{3}}(g_0)_{11},\biggl(\frac{\mu_1\mu_2}{\rho^2}\biggr)^{\frac{2}{3}}(g_0)_{22},\biggl(\frac{\mu_1\mu_2}{\rho^2}\biggr)^{-\frac{4}{3}}(g_0)_{33}\Bigg).
\end{eqnarray}
We find that 
the solution above is not regular on the part of the axis $z<-\sigma$ 
due to the existence of the factors 
$(\mu_1\mu_2/\rho^2)^{\frac{2}{3}},(\mu_1\mu_2/\rho^2)^{-\frac{4}{3}}$;  
they behave as $(\mu_1\mu_2/\rho^2)^{\frac{2}{3}}\sim \rho^{-\frac{4}{3}},(\mu_1\mu_2/\rho^2)^{-\frac{4}{3}}\sim\rho^{\frac{8}{3}}$ there. 
Since we use Minkowski spacetime as a seed in this paper, 
we cannot obtain a solution which is regular everywhere on the axis and 
has two angular momentum components. This is why we 
restrict ourselves to only solutions with a single angular momentum component. 
   
 However, we should note that there exists some freedom 
when we construct a physical metric with one 
angular momentum component from an unphysical metric with one angular momentum 
component by multiplying the normalization factors. 
In $e=f=0$ (or $c=d=0)$ case, the three dimensional unphysical metric 
$g^{{\rm (unphys)}}$ can be decomposed into the $2+1$ block matrix,

\begin{eqnarray}
g^{{\rm (unphys)}}=\left(
\begin{array}{@{\,}c|ccc@{\,}}
g^{{\rm (unphys)}}_{AB} & 0 \\ \hline
0                      & (g_0)_{33}\label{eq:21}

\end{array}
\right),
\end{eqnarray}
where $g^{{\rm (unphys)}}_{AB} (A,B=1,2)$ is a $2\times 2$ matrix 
dependent on the four parameters $a,b,c,d$. In this case, in order to 
satisfy the constraint ${\rm det} g=-\rho^2$, we may choose a normalization which multiplies $g^{{\rm (unphys)}}_{AB}$ by the normalization factor of four dimensions i.e. put $D=4$ and $n=2$ in the equation (\ref{eq:nor}). We leave the remaining component $(g_0)_{33}$ intact, i.e.

\begin{eqnarray}
g^{{\rm (phys)}}=\left(
\begin{array}{@{\,}c|ccc@{\,}}
-\frac{\mu_1\mu_2}{\rho^2}g^{{\rm (unphys)}}_{AB} & 0 \\ \hline
0                      & (g_0)_{33}\label{eq:211}

\end{array}
\right)\label{eq:physical}.
\end{eqnarray}
We can easily show that if a seed metric satisfies the condition
${\rm det} g_0=-\rho^2$, the physical metric (\ref{eq:physical}) also 
satisfies this condition. (In $c=d=0$ case, we can choose the same 
normalization which leaves $(g_0)_{22}$ intact and multiplies the 
$g_{\tilde A\tilde B}^{{\rm (unphys)}} (\tilde A,\tilde B=1,3)$ by 
$-\frac{\mu_1\mu_2}{\rho^2}$.)

\subsection{Static seed solutions}
The assumption that seed solutions $g_0$ are static simplifies all 
analysis since we can assume that the generating matrix $\psi_0$ becomes 
diagonal like $\psi_0={\rm diag} (\psi_1,\psi_2,\psi_3)$, where 
$\psi_a\ (a=1,2,3)$ are functions which depend on $\lambda,\ \rho$ and $z$. 
Then, the partial derivative equations for the generating matrix $\psi_0$ 
are decoupled into each component and therefore, we can solve each 
$\psi_a\ (a=1,2,3)$ independently. The unphysical metric~(\ref{eq:unphys}) 
becomes

\begin{eqnarray}
g^{{\rm (unphys)}}_{ab}=(g_0)_{ab}-(g_0)_{aa}(g_0)_{bb}\frac{\rho^2+\mu_1\mu_2}{\mu_1^2\mu_2^2}\frac{\Sigma_{ab}}{\Sigma'},\label{eq:un1}
\end{eqnarray}
where the functions $\Sigma'(\rho,z)$ and 
$\Sigma_{ab}(\rho,z)\ (a,b=1,2,3)$ are given by the equations~(\ref{eq:a})-(\ref{eq:a23}) in Appendix A. 
Since the expression of the metric is much lengthy, we do not write it here.

\section{solutions generated from Minkowski seed}
\label{sec:Minkowski_seed}
In four dimensional stationary and axisymmetric spacetimes, the solution generated from flat background as a seed is most interesting, since it was shown that a double soliton on a Minkowski background gives a Kerr solution, which is one of physically most important black hole solutions~\cite{Belinskii}. Therefore, we can expect that in five dimensions, this inverse scattering method might also give us interesting and important black holes with asymptotic flatness, and great insight into higher dimensional black holes. In this section, we focus on the simplest case, i.e. a two solitonic solution on five dimensional flat background spacetime. As a seed solution $g_{0}$, we choose Minkowski spacetime whose metric is given by~\cite{Harmark:2004rm}
\begin{eqnarray}
ds^2=-dt^2+\lambda_1d\phi^2+\lambda_2d\psi^2+\frac{1}{2\sqrt{\rho^2+(z+\kappa\sigma)^2}}(d\rho^2+dz^2),
\end{eqnarray}
where
\begin{eqnarray}
\lambda_1=\sqrt{\rho^2+(z+\kappa\sigma)^2}-(z+\kappa\sigma),\quad \lambda_2=\sqrt{\rho^2+(z+\kappa\sigma)^2}+(z+\kappa\sigma).\label{eq:lambda1}
\end{eqnarray}
In this paper we assume $\kappa$ is a parameter satisfying $\kappa\ge 1$. 
Here we put $x^1=t$, $x^2=\phi$, $x^3=\psi$, and then $g_0={\rm diag}(-1,\lambda_1,\lambda_2)$. 
The $3\times 3$ matrices $U_0$ and $V_0$ corresponding to $g_0$ are expressed in the form 
\begin{eqnarray}
U_0=\rho g_{0,\rho}g_0^{-1}={\rm diag}\Biggl(0,1+\frac{z+\kappa\sigma}{\sqrt{\rho^2+(z+\kappa\sigma)^2}},1-\frac{z+\kappa\sigma}{\sqrt{\rho^2+(z+\kappa\sigma)^2}}\Biggr),
\end{eqnarray}

\begin{eqnarray}
V_0=\rho g_{0,z}g_0^{-1}={\rm diag}\Biggl(0,-\frac{\rho}{\sqrt{\rho^2+(z+\kappa\sigma)^2}},\frac{\rho}{\sqrt{\rho^2+(z+\kappa\sigma)^2}}\Biggr).
\end{eqnarray}

The generating matrix $\psi_0$ for the static seed $g_0$
is diagonal, i.e. 
$\psi_0(\lambda,\rho,z)={\rm diag}(\psi_1(\lambda,\rho,z),\psi_2(\lambda,\rho,z),\psi_3(\lambda,\rho,z))$. 
Its components are given by

\begin{eqnarray}
& &\psi_1(\lambda,\rho,z)=-1,\nonumber\\
& &\psi_2(\lambda,\rho,z)=\lambda_1-\lambda=\sqrt{\rho^2+(z+\kappa\sigma)^2}-z-\kappa\sigma-\lambda,\nonumber\\
& &\psi_3(\lambda,\rho,z)=\lambda_2+\lambda=\sqrt{\rho^2+(z+\kappa\sigma)^2}+z+\kappa\sigma+\lambda\label{eq:minkowski}
\end{eqnarray}
with $\psi_0(\lambda=0,\rho,z)=g_0$.
Substituting the equations (\ref{eq:minkowski}), (\ref{eq:mu1}) and (\ref{eq:mu2}) into the equation (\ref{eq:un1}), we obtain the unphysical metric with eight parameters $\{\sigma,\kappa, a,b,c,d,e,f\}$.

\subsection{Single angular momentum case}
As mentioned in the previous section, we focus on 
a solution with a single angular momentum; this is the case where the 
three dimensional metric $g$ is block-diagonalized as the equations (\ref{eq:21}) and (\ref{eq:211}).  
Choosing the parameters as $e=f=0$, we study a physical metric which is generated with four dimensional normalization, as described in the equation (\ref{eq:physical}).  
The solution can be written in the following form:
\begin{eqnarray}
g^{{\rm (phys)}}_{11}=-\frac{G_{11}}{\mu_1\mu_2\Sigma},
\quad
g_{12}^{{\rm (phys)}}=-\lambda_1\frac{(\rho^2+\mu_1\mu_2)G_{12}}{\mu_1\mu_2 \Sigma},
\quad
g^{{\rm (phys)}}_{22}=-\lambda_1\frac{G_{22}}{\mu_1\mu_2\Sigma},
\label{eq:gphys}
\end{eqnarray}

\begin{eqnarray}
g^{{\rm (phys)}}_{33}=\lambda_2,
\quad
g_{23}^{{\rm (phys)}}=g_{13}^{{\rm (phys)}}=0,
\end{eqnarray}
where $\mu_1$, $\mu_2$, $\lambda_1$ and $\lambda_2$ are given by the 
equations~(\ref{eq:mu1}),~(\ref{eq:mu2}) and~(\ref{eq:lambda1}).
The functions $G_{11},\ G_{12},\ G_{22}$ and $\Sigma$ are defined as

\begin{eqnarray}
        G_{11}&=&-a^2b^2(\lambda_1-\mu_1)^2(\lambda_1-\mu_2)^2(\mu_1-\mu_2)^2\rho^4\nonumber\\
             &+&a^2d^2\lambda_1 \mu_2^2(\rho^2+\mu_1\mu_2)^2(\lambda_1-\mu_1)^2\nonumber\\
             &+&b^2c^2\lambda_1 \mu_1^2(\rho^2+\mu_1\mu_2)^2(\lambda_1-\mu_2)^2\nonumber\\
             &-&c^2d^2\lambda_1^2\mu_1^2\mu_2^2(\mu_1-\mu_2)^2\nonumber\\
             &-&2abcd\lambda_1(\lambda_1-\mu_1)(\lambda_1-\mu_2)(\rho^2+\mu_1^2)(\rho^2+\mu_2^2)\mu_1\mu_2,
\end{eqnarray}

\begin{eqnarray}
G_{22}&=&a^2b^2\mu_1^2\mu_2^2(\mu_1-\mu_2)^2(\lambda_1-\mu_1)^2(\lambda_1-\mu_2)^2\nonumber\\
      &+&c^2d^2\lambda_1^2(\mu_1-\mu_2)^2\rho^4\nonumber\\
      &-&a^2d^2\lambda_1\mu_1^2(\lambda_1-\mu_1)^2(\rho^2+\mu_1\mu_2)^2\nonumber\\
      &-&b^2c^2\lambda_1\mu_2^2(\lambda_1-\mu_2)^2(\rho^2+\mu_1\mu_2)^2\nonumber\\
      &+&2abcd\lambda_1\mu_1\mu_2(\lambda_1-\mu_2)(\lambda_1-\mu_1)(\rho^2+\mu_1^2)(\rho^2+\mu_2^2),
\end{eqnarray}

\begin{eqnarray}
G_{12}&=&ab^2c\mu_2(\mu_1-\mu_2)(\lambda_1-\mu_2)^2(\lambda_1-\mu_1)(\rho^2+\mu_1^2)\nonumber\\
&+&acd^2\lambda_1\mu_2(\mu_2-\mu_1)(\lambda_1-\mu_1)(\rho^2+\mu_1^2)\nonumber\\
&+&a^2bd\mu_1(\mu_2-\mu_1)(\lambda_1-\mu_1)^2(\lambda_1-\mu_2)(\rho^2+\mu_2^2)\nonumber\\
&+&bc^2d\mu_1\lambda_1(\lambda_1-\mu_2)(\rho^2+\mu_2^2)(\mu_1-\mu_2),
\end{eqnarray}

\begin{eqnarray}
\Sigma&=&a^2b^2(\lambda_1-\mu_1)^2(\lambda_1-\mu_2)^2(\mu_1-\mu_2)^2\rho^2+c^2d^2\lambda_1^2(\mu_1-\mu_2)^2\rho^2\nonumber\\
      & &+a^2d^2\lambda_1(\lambda_1-\mu_1)^2(\rho^2+\mu_1\mu_2)^2
      +c^2b^2\lambda_1(\lambda_1-\mu_2)^2(\rho^2+\mu_1\mu_2)^2\nonumber\\
      & &-2abcd\lambda_1(\lambda_1-\mu_1)(\lambda_1-\mu_2)(\rho^2+\mu_1^2)(\rho^2+\mu_2^2).
\end{eqnarray}
In order to see relation between the generated solution and the solution obtained by Mishima and Iguchi, let us consider the coordinate transformation of the physical metric such that
\begin{eqnarray}
t\rightarrow t'=t-\omega\phi, \quad \phi\rightarrow\phi'=\phi,
\end{eqnarray}
where $\omega$ is an arbitrary constant and $x^1=t, x^2=\phi,x^3=\psi$. 
We should note that the transformed metric also satisfies the 
supplementary condition ${\rm det} g=-\rho^2$.
Under this transformation, the physical metric components become
\begin{eqnarray}
& & g_{tt}\rightarrow g_{t't'}=g_{tt},\label{eq:trtt}\\ 
& &g_{t\phi}\rightarrow g_{t'\phi'}=g_{t\phi}+\omega g_{tt},\label{eq:trtp}\\ 
& &g_{\phi\phi}\rightarrow g_{\phi'\phi'}=g_{\phi\phi}+2\omega g_{t\phi}+\omega^2g_{tt}.\label{eq:trpp}
\end{eqnarray}
If we choose the parameters such that 
\begin{eqnarray}
ab&=&\beta,\\
bc&=&\sigma^{\frac{1}{2}}(\kappa-1),\\
ad&=&-\sigma^{\frac{1}{2}}\alpha\beta(\kappa+1),\\
cd&=&-\sigma\alpha(\kappa^2-1),\\
\omega&=&C_1,
\end{eqnarray}
and use the spherical polar coordinate $(x,y)$ defined as 
$\rho=\sigma\sqrt{(x^2-1)(1-y^2)},\ z=\sigma xy$, then we can make sure 
that the transformed metric with components 
$(\ref{eq:trtt}),(\ref{eq:trtp})$ and $(\ref{eq:trpp})$  exactly coincide with 
the metric (the equations (\ref{eq:pp}) and (\ref{eq:MI}) in Appendix C) of a black ring solution with a rotating two sphere~\cite{Mishima:2005id,Pomeransky:2005sj}. We should note that our solution coincides with the original expression of the metric derived by Mishima and Iguchi( eq.(6) in ref. \cite{Mishima:2005id}) before we choose such parameters that closed time like curves vanises~\cite{Private}. 
In order to show this coincidence, it is sufficient to calculate only two 
components $g_{tt}$ and $g_{t\phi}$ due to the supplementary condition 
${\rm det} g=-\rho^2$ and the fact that the metric function $f$ is 
determined by three dimensional metric $g$.

\section{Summary and discussion}
\label{sec:summary}
In this article, we studied two solitonic solutions of vacuum 
Einstein equations from five dimensional Minkowski spacetime using the 
inverse scattering method.
The solution with one angular momentum includes six parameters $\{a,b,c,d,\sigma,\kappa\}$, however, physical parameters are only four since the only ratios of $a/c$ and $b/d$ appear in the metric components~(\ref{eq:gphys}) i.e. the transformations of parameters which leave these ratios invariant are isometries.   
As a result, we reproduced a black ring 
solution with a rotating two sphere which was obtained by Mishima and 
Iguchi.
However, we could not obtain a black ring solution which was found by 
Emparan and Reall from Minkowski spacetime with this inverse 
scattering method, though we tried to find that solution by restricting the parameters to $c=d=0$.

We also discussed the possibility that one can derive a black ring 
solution with two angular momentum components from Minkowski seed 
solution. We found that the generated solution cannot be regular on the 
part of an axis if the seed solution is regular there. Therefore, in 
order to obtain regular and asymptotically flat black ring/hole solutions 
by this inverse scattering method, we may need a singular seed solution.
In this stage, we do not know how to generate such solutions.

\section*{Acknowledgements}
We thank Masato~Nozawa for useful comments.
We would like to thank Takashi~Mishima and Hideo~Iguchi 
for continuous discussion. We also thank Ken-ichi~Nakao and  
Hideki~Ishihara for continuous encouragement.
This work is supported by the 21 COE program ``Constitution of 
wide-angle mathematical basis focused on knots.'' The work of Y.~Y. is 
supported by the Grant-in Aid for Scientific Research (No.~17540262 and 
No.~17540091) from Japan Ministry of Education.

\section*{Note added}
After we submitted this article to e-prints archives, we noticed a new article~\cite{Azuma}, which has a considerable overlap with our article. In \cite{Azuma}, the Myers-Perry solution with one angular momentum was reproduced by the inverse scattering method. However, the black ring solution with a rotating two sphere, which we reproduced in this article by the inverse scattering method, contains the Myers-Perry black hole solution with a single angular momentum. Therefore, $(2,0)$ soliton solution in ref. \cite{Azuma} are contained in our result, that is, if in our article we put $\kappa=1$ in eq. (\ref{eq:lambda1}), our result exactly coincides with their solution.

\section{Appendix}
\subsection{General solutions generated from static seeds}
In this section, we describe general solutions generated from static 
seed solutions, where we may assume the generating matrix 
$\psi_0(\lambda,\rho,z)$ to be diagonal, i.e.
$\psi_0={\rm diag}(\psi_1,\psi_2,\psi_3)$.
The unphysical metric can be written in the considerably long expression ;  

\begin{eqnarray}
g^{{\rm (unphys)}}_{ab}=(g_0)_{ab}-(g_0)_{aa}(g_0)_{bb}\frac{\rho^2+\mu_1\mu_2}{\mu_1^2\mu_2^2}\frac{\Sigma_{ab}}{\Sigma'}\label{eq:un},
\end{eqnarray}
where the functions $\Sigma'(\rho,z)$, $\Sigma_{ab}(\rho,z)\ (a,b=1,2,3)$ are defined as
\begin{eqnarray}
\Sigma'&=&-a^2b^2\psi_2(\mu_1)^2\psi_2(\mu_2)^2\psi_3(\mu_1)^2\psi_3(\mu_2)^2(g_0)_{11}^2(\mu_1-\mu_2)^2\rho^2\nonumber\\
      &-&c^2d^2\psi_1(\mu_1)^2\psi_1(\mu_2)^2\psi_3(\mu_1)^2\psi_3(\mu_2)^2(g_0)_{22}^2(\mu_1-\mu_2)^2\rho^2\nonumber\\
      &-&e^2f^2\psi_1(\mu_1)^2\psi_1(\mu_2)^2\psi_2(\mu_1)^2\psi_2(\mu_2)^2 (g_0)_{33}^2(\mu_1-\mu_2)^2\rho^2\nonumber\\
      &+&a^2d^2\psi_1(\mu_2)^2\psi_2(\mu_1)^2\psi_3(\mu_1)^2\psi_3(\mu_2)^2(g_0)_{11}(g_0)_{22}(\rho^2+\mu_1\mu_2)^2\nonumber\\
      &+&a^2f^2\psi_1(\mu_2)^2\psi_2(\mu_1)^2\psi_2(\mu_2)^2\psi_3(\mu_1)^2(g_0)_{11}(g_0)_{33}(\rho^2+\mu_1\mu_2)^2\nonumber\\
      &+&b^2c^2\psi_1(\mu_1)^2\psi_2(\mu_2)^2\psi_3(\mu_1)^2\psi_3(\mu_2)^2(g_0)_{11}(g_0)_{22}(\rho^2+\mu_1\mu_2)^2\nonumber\\
      &+&c^2f^2\psi_1(\mu_1)^2\psi_1(\mu_2)^2\psi_2(\mu_2)^2\psi_3(\mu_1)^2(g_0)_{22}(g_0)_{33}(\rho^2+\mu_1\mu_2)^2\nonumber\\
      &+&b^2e^2\psi_1(\mu_1)^2\psi_2(\mu_1)^2\psi_2(\mu_2)^2\psi_3(\mu_2)^2(g_0)_{11}(g_0)_{33}(\rho^2+\mu_1\mu_2)^2\nonumber\\
      &+&d^2e^2\psi_1(\mu_1)^2\psi_1(\mu_2)^2\psi_2(\mu_1)^2\psi_3(\mu_2)^2(g_0)_{22}(g_0)_{33}(\rho^2+\mu_1\mu_2)^2\nonumber\\
      &-&2abcd\psi_1(\mu_1)\psi_1(\mu_2)\psi_2(\mu_1)\psi_2(\mu_2)\psi_3(\mu_1)^2\psi_3(\mu_2)^2 (g_0)_{11}(g_0)_{22}(\rho^2+\mu_1^2)(\rho^2+\mu_2^2)\nonumber\\
      &-&2abef\psi_1(\mu_1)\psi_1(\mu_2)\psi_2(\mu_1)^2\psi_2(\mu_2)^2\psi_3(\mu_1)\psi_3(\mu_2) (g_0)_{11}(g_0)_{33}(\rho^2+\mu_1^2)(\rho^2+\mu_2^2)\nonumber\\
      &-&2cdef\psi_1(\mu_1)^2\psi_1(\mu_2)^2\psi_2(\mu_1)\psi_2(\mu_2)\psi_3(\mu_1)\psi_3(\mu_2) (g_0)_{22}(g_0)_{33}(\rho^2+\mu_1^2)(\rho^2+\mu_2^2),\label{eq:a}
\end{eqnarray}

\begin{eqnarray}
\Sigma_{11}&=&a^2b^2\psi_2(\mu_1)^2\psi_2(\mu_2)^2\psi_3(\mu_1)^2\psi_3(\mu_2)^2(g_0)_{11}(\mu_1-\mu_2)^2\rho^2(\rho^2-\mu_1\mu_2)\nonumber\\
           &+&a^2d^2\psi_1(\mu_2)^2\psi_2(\mu_1)^2\psi_3(\mu_1)^2\psi_3(\mu_2)^2(g_0)_{22}\mu_2^2(\rho^2+\mu_1^2)(\rho^2+\mu_1\mu_2)\nonumber\\
           &+&a^2f^2\psi_1(\mu_2)^2\psi_2(\mu_1)^2\psi_2(\mu_2)^2\psi_3(\mu_1)^2(g_0)_{33}\mu_2^2(\rho^2+\mu_1^2)(\rho^2+\mu_1\mu_2)\nonumber\\
           &+&b^2c^2\psi_1(\mu_1)^2\psi_2(\mu_2)^2\psi_3(\mu_1)^2\psi_3(\mu_2)^2(g_0)_{22}\mu_1^2(\rho^2+\mu_2^2)(\rho^2+\mu_1\mu_2)\nonumber\\
           &+&b^2e^2\psi_1(\mu_1)^2\psi_2(\mu_1)^2\psi_2(\mu_2)^2\psi_3(\mu_2)^2(g_0)_{33}\mu_1^2(\rho^2+\mu_2^2)(\rho^2+\mu_1\mu_2)\nonumber\\
           &-&2abcd\psi_1(\mu_1)\psi_1(\mu_2)\psi_2(\mu_1)\psi_2(\mu_2)\psi_3(\mu_1)^2\psi_3(\mu_2)^2 (g_0)_{22}\mu_1\mu_2(\rho^2+\mu_1^2)(\rho^2+\mu_2^2)\nonumber\\
           &-&2abef\psi_1(\mu_1)\psi_1(\mu_2)\psi_2(\mu_1)^2\psi_2(\mu_2)^2\psi_3(\mu_1)\psi_3(\mu_2) (g_0)_{33}\mu_1\mu_2(\rho^2+\mu_1^2)(\rho^2+\mu_2^2),\label{eq:a11}
\end{eqnarray}

\begin{eqnarray}
\Sigma_{22}&=&a^2d^2\psi_1(\mu_2)^2\psi_2(\mu_1)^2\psi_3(\mu_1)^2\psi_3(\mu_2)^2(g_0)_{11}\mu_1^2(\rho^2+\mu_2^2)(\rho^2+\mu_1\mu_2)\nonumber\\
          &+&b^2c^2\psi_1(\mu_1)^2\psi_2(\mu_2)^2\psi_3(\mu_1)^2\psi_3(\mu_2)^2(g_0)_{11}\mu_2^2(\rho^2+\mu_1^2)(\rho^2+\mu_1\mu_2)\nonumber\\
          &+&c^2d^2\psi_1(\mu_1)^2\psi_1(\mu_2)^2\psi_3(\mu_1)^2\psi_3(\mu_2)^2(g_0)_{22}(\mu_1-\mu_2)^2\rho^2(\rho^2-\mu_1\mu_2)\nonumber\\
          &+&c^2f^2\psi_1(\mu_1)^2\psi_1(\mu_2)^2\psi_2(\mu_2)^2\psi_3(\mu_1)^2(g_0)_{33}\mu_2^2(\rho^2+\mu_1^2)(\rho^2+\mu_1\mu_2)\nonumber\\
          &+&d^2e^2\psi_1(\mu_1)^2\psi_1(\mu_2)^2\psi_2(\mu_1)^2\psi_3(\mu_2)^2(g_0)_{33}\mu_1^2(\rho^2+\mu_2^2)(\rho^2+\mu_1\mu_2)\nonumber\\
           &-&2abcd\psi_1(\mu_1)\psi_1(\mu_2)\psi_2(\mu_1)\psi_2(\mu_2)\psi_3(\mu_1)^2\psi_3(\mu_2)^2 (g_0)_{22}\mu_1\mu_2(\rho^2+\mu_1^2)(\rho^2+\mu_2^2)\nonumber\\
           &-&2cdef\psi_1(\mu_1)^2\psi_1(\mu_2)^2\psi_2(\mu_1)\psi_2(\mu_2)\psi_3(\mu_1)\psi_3(\mu_2) (g_0)_{33}\mu_1\mu_2(\rho^2+\mu_1^2)(\rho^2+\mu_2^2),\label{eq:a22}
\end{eqnarray}

\begin{eqnarray}
\Sigma_{33}&=&a^2f^2\psi_1(\mu_2)^2\psi_2(\mu_1)^2\psi_2(\mu_2)^2\psi_3(\mu_1)^2(g_0)_{11}\mu_1^2(\rho^2+\mu_2^2)(\rho^2+\mu_1\mu_2)\nonumber\\
           &+&b^2e^2\psi_1(\mu_1)^2\psi_2(\mu_1)^2\psi_2(\mu_2)^2\psi_3(\mu_2)^2(g_0)_{11}\mu_2^2(\rho^2+\mu_1^2)(\rho^2+\mu_1\mu_2)\nonumber\\
            &+&c^2f^2\psi_1(\mu_1)^2\psi_1(\mu_2)^2\psi_2(\mu_2)^2\psi_3(\mu_1)^2(g_0)_{22}\mu_1^2(\rho^2+\mu_2^2)(\rho^2+\mu_1\mu_2)\nonumber\\
            &+&d^2e^2\psi_1(\mu_1)^2\psi_1(\mu_2)^2\psi_2(\mu_1)^2\psi_3(\mu_2)^2(g_0)_{22}\mu_2^2(\rho^2+\mu_1^2)(\rho^2+\mu_1\mu_2)\nonumber\\
            &+&e^2f^2\psi_1(\mu_1)^2\psi_1(\mu_2)^2\psi_2(\mu_1)^2\psi_2(\mu_2)^2 (g_0)_{33}(\mu_1-\mu_2)^2\rho^2(\rho^2-\mu_1\mu_2)\nonumber\\
            &-&2abef\psi_1(\mu_1)\psi_1(\mu_2)\psi_2(\mu_1)^2\psi_2(\mu_2)^2\psi_3(\mu_1)\psi_3(\mu_2) (g_0)_{11}\mu_1\mu_2(\rho^2+\mu_1^2)(\rho^2+\mu_2^2)\nonumber\\
            &-&2cdef\psi_1(\mu_1)^2\psi_1(\mu_2)^2\psi_2(\mu_1)\psi_2(\mu_2)\psi_3(\mu_1)\psi_3(\mu_2) (g_0)_{22}\mu_1\mu_2(\rho^2+\mu_1^2)(\rho^2+\mu_2^2),\label{eq:a33}
\end{eqnarray}

\begin{eqnarray}
\Sigma_{12}&=&ab^2c\psi_1(\mu_1)\psi_2(\mu_1)\psi_2(\mu_2)^2\psi_3(\mu_1)^2\psi_3(\mu_2)^2(g_0)_{11}\mu_2(\mu_2-\mu_1)(\rho^2+\mu_1^2)\rho^2\nonumber\\
           &+&acd^2\psi_1(\mu_1)\psi_1(\mu_2)^2\psi_2(\mu_1)\psi_3(\mu_1)^2\psi_3(\mu_2)^2(g_0)_{22}\mu_2(\mu_2-\mu_1)(\rho^2+\mu_1^2)\rho^2\nonumber\\
           &+&a^2bd\psi_1(\mu_2)\psi_2(\mu_1)^2\psi_2(\mu_2)\psi_3(\mu_1)^2\psi_3(\mu_2)^2(g_0)_{11}\mu_1(\mu_1-\mu_2)(\rho^2+\mu_2^2)\rho^2\nonumber\\
           &+&bc^2d\psi_1(\mu_1)^2\psi_1(\mu_2)\psi_2(\mu_2)\psi_3(\mu_1)^2\psi_3(\mu_2)^2(g_0)_{22}\mu_1(\mu_1-\mu_2)(\rho^2+\mu_2^2)\rho^2\nonumber\\
           &+&acf^2\psi_1(\mu_1)\psi_1(\mu_2)^2\psi_2(\mu_1)\psi_2(\mu_2)^2\psi_3(\mu_1)^2(g_0)_{33}\mu_2^2(\rho^2+\mu_1^2)(\rho^2+\mu_1\mu_2)\nonumber\\
           &+&bde^2\psi_1(\mu_1)^2\psi_1(\mu_2)\psi_2(\mu_1)^2\psi_2(\mu_2)\psi_3(\mu_2)^2(g_0)_{33}\mu_1^2(\rho^2+\mu_2^2)(\rho^2+\mu_1\mu_2)\nonumber\\
           &-&adef\psi_1(\mu_1)\psi_1(\mu_2)^2\psi_2(\mu_1)^2\psi_2(\mu_2)\psi_3(\mu_1)\psi_3(\mu_2) (g_0)_{33}\mu_1\mu_2(\rho^2+\mu_1^2)(\rho^2+\mu_2^2)\nonumber\\
           &-&bcef\psi_1(\mu_1)^2\psi_1(\mu_2)\psi_2(\mu_1)\psi_2(\mu_2)^2\psi_3(\mu_1)\psi_3(\mu_2) (g_0)_{33}\mu_1\mu_2(\rho^2+\mu_1^2)(\rho^2+\mu_2^2),\label{eq:a12}
\end{eqnarray}

\begin{eqnarray}
\Sigma_{13}&=&ab^2e\psi_1(\mu_1)\psi_2(\mu_1)^2\psi_2(\mu_2)^2\psi_3(\mu_1)\psi_3(\mu_2)^2(g_0)_{11}\mu_2(\mu_2-\mu_1)(\rho^2+\mu_1^2)\rho^2\nonumber\\
           &+&aef^2\psi_1(\mu_1)\psi_1(\mu_2)^2\psi_2(\mu_1)^2\psi_2(\mu_2)^2\psi_3(\mu_1) (g_0)_{33}\mu_2(\mu_2-\mu_1)(\rho^2+\mu_1^2)\rho^2\nonumber\\
           &+&a^2bf\psi_1(\mu_2)\psi_2(\mu_1)^2\psi_2(\mu_2)^2\psi_3(\mu_1)^2\psi_3(\mu_2) (g_0)_{11}\mu_1(\mu_1-\mu_2)(\rho^2+\mu_2^2)\rho^2\nonumber\\
           &+&be^2f\psi_1(\mu_1)^2\psi_1(\mu_2)\psi_2(\mu_1)^2\psi_2(\mu_2)^2\psi_3(\mu_2) (g_0)_{33}\mu_1(\mu_1-\mu_2)(\rho^2+\mu_2^2)\rho^2\nonumber\\
           &+&ad^2e\psi_1(\mu_1)\psi_1(\mu_2)^2\psi_2(\mu_1)^2\psi_3(\mu_1)\psi_3(\mu_2)^2 (g_0)_{22}\mu_2^2(\rho^2+\mu_1^2)(\rho^2+\mu_1\mu_2)\nonumber\\
           &+&bc^2f\psi_1(\mu_1)^2\psi_1(\mu_2)\psi_2(\mu_1)^2\psi_3(\mu_1)^2\psi_3(\mu_2) (g_0)_{22}\mu_1^2(\rho^2+\mu_2^2)(\rho^2+\mu_1\mu_2)\nonumber\\
           &-&acdf\psi_1(\mu_1)\psi_1(\mu_2)^2\psi_2(\mu_1)\psi_2(\mu_2)\psi_3(\mu_1)^2\psi_3(\mu_2) (g_0)_{22}\mu_1\mu_2(\rho^2+\mu_1^2)(\rho^2+\mu_2^2)\nonumber\\
           &-&bcde\psi_1(\mu_1)^2\psi_1(\mu_2)\psi_2(\mu_1)\psi_2(\mu_2)\psi_3(\mu_1)\psi_3(\mu_2)^2 (g_0)_{22}\mu_1\mu_2(\rho^2+\mu_1^2)(\rho^2+\mu_2^2),\label{eq:a13}
\end{eqnarray}

\begin{eqnarray}
\Sigma_{23}&=&cd^2e\psi_1(\mu_1)^2\psi_1(\mu_2)^2\psi_2(\mu_1)\psi_3(\mu_1)\psi_3(\mu_2)^2 (g_0)_{22}\mu_2(\mu_2-\mu_1)(\rho^2+\mu_1^2)(\rho^2+\mu_2^2)\nonumber\\
           &+&cef^2\psi_1(\mu_1)^2\psi_1(\mu_2)^2\psi_2(\mu_1)\psi_2(\mu_2)^2\psi_3(\mu_1) (g_0)_{33}\mu_2(\mu_2-\mu_1)(\rho^2+\mu_1^2)(\rho^2+\mu_2^2)\nonumber\\
           &+&c^2df\psi_1(\mu_1)^2\psi_1(\mu_2)^2\psi_2(\mu_2)\psi_3(\mu_1)^2\psi_3(\mu_2) (g_0)_{22}\mu_1(\mu_1-\mu_2)(\rho^2+\mu_1^2)(\rho^2+\mu_2^2)\nonumber\\
           &+&de^2f\psi_1(\mu_1)^2\psi_1(\mu_1)^2\psi_2(\mu_1)^2\psi_2(\mu_2)\psi_3(\mu_2) (g_0)_{33}\mu_1(\mu_1-\mu_2)(\rho^2+\mu_1^2)(\rho^2+\mu_2^2)\nonumber\\
           &+&b^2ce\psi_1(\mu_1)^2\psi_2(\mu_1)^2\psi_2(\mu_2)^2\psi_3(\mu_1)\psi_3(\mu_2)^2 (g_0)_{11}\mu_2^2(\rho^2+\mu_1^2)(\rho^2+\mu_1\mu_2)\nonumber\\
           &+&a^2df\psi_1(\mu_2)^2\psi_2(\mu_1)^2\psi_2(\mu_2)\psi_3(\mu_1)^2\psi_3(\mu_2) (g_0)_{11}\mu_1^2(\rho^2+\mu_2^2)(\rho^2+\mu_1\mu_2)\nonumber\\
           &-&abcf\psi_1(\mu_1)\psi_1(\mu_2)\psi_2(\mu_1)\psi_2(\mu_2)^2\psi_3(\mu_1)\psi_3(\mu_2)^2\mu_1\mu_2(\rho^2+\mu_1^2)(\rho^2+\mu_2^2)\nonumber\\
           &-&abde\psi_1(\mu_1)\psi_1(\mu_2)\psi_2(\mu_1)^2\psi_2(\mu_2)\psi_3(\mu_1)\psi_3(\mu_2)^2\mu_1\mu_2(\rho^2+\mu_1^2)(\rho^2+\mu_2^2).\label{eq:a23}
\end{eqnarray}

\subsection{Conformal rescale}

Here we study how the generating matrix $\psi$ transforms under the transformation of the seed metric. If we perform the conformal transformation $g\rightarrow \tilde g=\Omega g$ for $\Omega$ such that $\Omega=\Omega(\rho,z)$ is a function dependent on $\rho$, $z$ and $\ln \Omega(\rho,z)$ is a harmonic function which is the solution of

\begin{eqnarray}
\triangle \ln \Omega\equiv \Biggl(\frac{\partial^2}{\partial\rho^2}+\frac{1}{\rho}\frac{\partial}{\partial \rho}+\frac{\partial^2}{\partial z^2}  \Biggr)\ln  \Omega=0,\label{eq:harmo}
\end{eqnarray}
then the generating matrix $\tilde \psi$ corresponding to $\tilde g$ satisfies

\begin{eqnarray}
D_1 (\ln\tilde\psi)&=&\frac{\rho V_0-\lambda U_0}{\rho^2+\lambda^2}+\frac{\rho^2\Omega_{,z}-\rho\lambda \Omega_{,\rho}}{\rho^2+\lambda^2},\label{eq:con1}
\end{eqnarray}
and

\begin{eqnarray}
D_2 (\ln\tilde\psi)&=&\frac{\rho U_0+\lambda V_0}{\rho^2+\lambda^2}+\frac{\rho^2\Omega_{,\rho}+\rho\lambda \Omega_{,z}}{\rho^2+\lambda^2}.\label{eq:con2}
\end{eqnarray}
Since these equations are linear, we easily find that the solution of the equations (\ref{eq:con1}) and (\ref{eq:con2}) can be expressed in the form of $\tilde \psi=\psi[\Omega]\cdot\psi$, where $\psi$ is the solution of equations (\ref{eq:psi1}) and (\ref{eq:psi2}) for the metric $g$, and  $\psi[\Omega]$ is the solution of the equations :
\begin{eqnarray}
D_1(\ln\psi[\Omega])=\frac{\rho^2\Omega_{,z}-\rho\lambda \Omega_\rho}{\rho^2+\lambda^2},\quad D_2(\ln\psi[\Omega])=\frac{\rho^2 \Omega_{,\rho}+\rho\lambda \Omega_{,z}}{\rho^2+\lambda^2}.\label{eq:con3}
\end{eqnarray}

If $\psi[\Omega_1]$ and $\psi[\Omega_2]$ are the solutions of (\ref{eq:con3}) for harmonic functions $\Omega_1$ and $\Omega_2$, respectively,
then we also find that the solution $\psi[\Omega_1\Omega_2]$ of (\ref{eq:con3}) for the function $\Omega_1\Omega_2$ satisfies the relation $\psi[\Omega_1\Omega_2]=\psi[\Omega_1]\psi[\Omega_2]$. We list the important solutions of the equations (\ref{eq:con3}) in our discussion,

\begin{eqnarray}
\psi[1]&=&1,\nonumber\\
\psi[\mu_1]&=&\mu_1-\lambda,\nonumber\\
\psi[\mu_2]&=&\mu_2-\lambda,\nonumber\\
\psi[\lambda_1]&=&\lambda_1-\lambda,\nonumber\\
\psi[\lambda_2]&=&\lambda_2+\lambda,\nonumber\\
\psi[\rho^2]&=&\rho^2-2z\lambda-\lambda^2.
\end{eqnarray}
where $\lambda_1$ and $\lambda_2$ are given by the equations~(\ref{eq:lambda1}).
From these results, the $\psi$ corresponding to the seed solution $g=\mu_1^a\mu_2^b\cdots\lambda_1^c\lambda_2^d\rho^{2e}g_0$ can be expressed in the form

\begin{eqnarray}
\psi=(\psi[\mu_1])^a(\psi[\mu_2])^b\cdots(\psi[\lambda_1])^c(\psi[\lambda_2])^d(\psi[\rho^2])^e\psi_0.
\end{eqnarray}

\subsection{Black ring solution with a rotating two sphere}
We describe the metric of a black ring solution with a rotating two 
sphere which was found by Mishima and Iguchi. Here we use the spherical 
polar coordinate $(x,y)$ defined by the equations 
$\rho=\sigma\sqrt{(x^2-1)(1-y^2)}$ and $z=\sigma xy$, as used in the Ref.~\cite{Mishima:2005id}. The explicit expression is

\begin{equation}
g_{tt}=-\frac{(x^2-1)(1+FG)^2-(1-y^2)(F-G)^2}{[(x+1)+(x-1)FG]^2+[(1+y)F+(1-y)G]^2},\label{eq:tt}
\end{equation}

\begin{eqnarray}
g_{t\phi}&=&\frac{2\sigma \lambda_1^{\frac{1}{2}}}{\rho}\frac{(x^2-1)(1+FG)[(1-y)G-(1+y)F]+(1-y^2)(G-F)[x+1+(1-x)FG]}{[(x+1)+(x-1)FG]^2+[(1+y)F+(1-y)G]^2}\nonumber\\
         & &+C_1\frac{(x^2-1)(1+FG)^2-(1-y^2)(F-G)^2}{[(x+1)-(x-1)FG]^2+[(1+y)F+(1-y)G]^2},\label{eq:tp}
\end{eqnarray}

\begin{equation}
g_{\phi\phi}=\frac{-\rho^2\lambda_2^{-1}+g_{t\phi}^2}{g_{tt}},\label{eq:pp}
\end{equation}
where $\lambda_1$ and $\lambda_2$ are given by the 
equations~(\ref{eq:lambda1}). 
Here the functions $F$, $G$ and the constant $C_1$ are defined as
\begin{eqnarray}
F=\alpha\frac{x-y+\kappa+1+\sqrt{x^2+y^2+2\kappa xy+\kappa^2-1}}{2(xy+\kappa+\sqrt{x^2+y^2+2\kappa xy+\kappa^2-1})^{\frac{1}{2}}},
\end{eqnarray}
\begin{eqnarray}
G=\beta\frac{2(xy+\kappa+\sqrt{x^2+y^2+2\kappa xy+\kappa^2-1})^{\frac{1}{2}}}{x+y+\kappa-1+\sqrt{x^2+y^2+2\kappa xy+\kappa^2-1}},
\end{eqnarray}
\begin{eqnarray}
C_1=\frac{2\sigma^{\frac{1}{2}}\alpha}{1+\alpha\beta}.\label{eq:c1}
\end{eqnarray}
Substituting the following equations into (\ref{eq:tt}), (\ref{eq:tp}) and (\ref{eq:pp}), 
\begin{eqnarray}
F=-\frac{\alpha(\kappa+1)\sigma^{\frac{3}{2}}\lambda_1^{\frac{1}{2}}}{\rho}\frac{(x+1)(1-y)}{\psi_2(\mu_2)},\quad 
G=-\frac{\beta\rho}{(\kappa-1)\sigma^{\frac{3}{2}}\lambda_1^{\frac{1}{2}}}\frac{\psi_2(\mu_1)}{(x-1)(1-y)},
\end{eqnarray}
we obtain the other expression of metric components,
\begin{eqnarray}
g_{tt}=-\frac{\tilde A}{\tilde B}, \quad g_{t\phi}=2\sigma^{\frac{1}{2}}\lambda_1\frac{\tilde C}{\tilde B}+C_1\frac{\tilde A}{\tilde B},\label{eq:MI}
\end{eqnarray}
where the functions $\tilde A,\ \tilde B$ and $\tilde C$ are defined as
\begin{eqnarray}
\tilde A&=&-\beta^2\psi_2(\mu_1)^2\psi_2(\mu_2)^2(1+y)^2+\sigma\alpha^2\beta^2(\kappa+1)^2\lambda_1\psi_2(\mu_1)^2(x+1)^2+\sigma(\kappa-1)^2\lambda_1\psi_2(\mu_2)^2(x-1)^2\nonumber\\
      &-&\sigma^2\alpha^2(\kappa^2-1)^2\lambda_1^2(1-y)^2+2\sigma\alpha\beta(\kappa^2-1)\lambda_1\psi_2(\mu_1)\psi_2(\mu_2)(x^2-y^2),
\end{eqnarray}

\begin{eqnarray}
\tilde B&=&\beta^2\psi_2(\mu_1)^2\psi_2(\mu_2)^2(1-y)^2+\sigma^2\alpha^2(\kappa^2-1)^2\lambda_1^2(1-y^2)+\sigma\alpha^2\beta^2(\kappa+1)^2\psi_2(\mu_1)^2\lambda_1(x^2-1)\nonumber\\
    &+&\sigma(\kappa-1)^2\psi_2(\mu_2)^2\lambda_1(x^2-1)+2\sigma\alpha\beta(\kappa^2-1)\lambda_1\psi_2(\mu_1)\psi_2(\mu_2)(x^2-y^2),
\end{eqnarray}

\begin{eqnarray}
\tilde C&=&-\beta(\kappa-1)\psi_2(\mu_1)\psi_2(\mu_2)^2(x+y)-\alpha\beta^2(\kappa+1)\psi_2(\mu_1)^2\psi_2(\mu_2)(x-y)\nonumber\\
    &+&\sigma\alpha^2\beta(\kappa^2-1)(\kappa+1)\lambda_1\psi_2(\mu_1)(x+y)+\sigma\alpha(\kappa^2-1)(\kappa-1)\psi_2(\mu_2)(x-y).
\end{eqnarray}


\begin{thebibliography}{22}
\bibitem{Private}
Mishima and Iguchi showed that the metric in the ref. \cite{Figueras} is identical to that in the ref. \cite{Mishima:2005id} if we restrict ourselves to the cases where closed timelike curves vanish. We knew this fact in the private discussion with them.
\end{thebibliography}
\end{document}